\documentclass[aps,prd,twocolumn,showpacs,floats,floatfix,letterpaper,nofootinbib,superscriptaddress,]{revtex4}

\usepackage{amssymb,amsmath,latexsym,mathrsfs}
\usepackage{graphicx}
\usepackage{epsfig}

\newcommand{\be}{\begin{equation}}
\newcommand{\e}{\end{equation}}
\newcommand{\bear}{\begin{eqnarray}}
\newcommand{\ear}{\end{eqnarray}}

\newcommand{\f}{\frac}
\newcommand{\de}{{\rm d}}

\newcommand{\xez}{$x_e(z)$\ }

\begin{document}


\title{Data-constrained reionization and its effects on cosmological parameters}

\author{S. Pandolfi}
\affiliation{Dark  Cosmology  Centre,  Niels  Bohr  Institute,  University  of  Copenhagen,  Juliane  Maries  Vej  30,  DK-2100  Copenhagen,
Denmark.}
\affiliation{Physics Department and ICRA, Universita' di Roma 
	``La Sapienza'', Ple.\ Aldo Moro 2, 00185, Rome, Italy}
\affiliation{Physics Department and INFN, Universita' di Roma 
	``La Sapienza'', Ple.\ Aldo Moro 2, 00185, Rome, Italy}
\author{A. Ferrara}
\affiliation{Scuola Normale Superiore, Piazza dei Cavalieri 7, 56126 Pisa, Italy}
\author{T. Roy Choudhury}
\affiliation{Harish-Chandra Research Institute, Chhatnag Road, Jhusi, Allahabad 211019 India}
\author{A. Melchiorri$^3$}
\noaffiliation
\author{S. Mitra$^5$}
\noaffiliation

\begin{abstract}
We perform an analysis of the recent WMAP7 data considering physically motivated and viable reionization
scenarios with the aim of assessing their effects on cosmological parameter determinations.
The main novelties are:
(i) the combination of CMB data with astrophysical results from quasar absorption line experiments; (ii)
the joint variation of both the cosmological and astrophysical [governing the evolution of the free electron fraction \xez]
parameters. Including a realistic, data-constrained reionization history in the analysis induces appreciable changes in the
cosmological parameter values deduced through a standard WMAP7 analysis.
Particularly noteworthy are the 
variations in  $\Omega_bh^2 = 0.02258^{+0.00057}_{-0.00056}$ (WMAP7)   vs.  $\Omega_bh^2 = 0.02183\pm 0.00054$ (WMAP7 + ASTRO),
and the new constraints for the scalar spectral index, for which WMAP7 + ASTRO excludes the Harrison-Zel'dovich value $n_s=1$ at $>3\sigma$.
Finally, the e.s. optical depth value is considerably decreased with respect to the standard WMAP7, i.e. $\tau_{e}=0.080\pm0.012$.
We conclude that the inclusion of astrophysical datasets, allowing to robustly constrain the reionization history, in the extraction procedure 
of cosmological parameters leads to relatively important differences in the final determination of their values. 
\end{abstract}

\maketitle
\section{Introduction}
It is well known from a large set of astrophysical observables that after primordial recombination (which occurred at a redshift of $z\sim1100$) the universe ``reionized'' at a redshift $z > 6$. It is common practice in Cosmic Microwave Background (CMB) studies to parametrize the reionization as an instantaneous process occurring at some redshift $z_r$, with $4<z_{r}<32$, and to marginalize over $z_r$ when deriving constraints on the other cosmological parameters. In the absence of any precise astrophysical model of the reionization process, the electron ionization fraction $x_e(z)$ is parametrized by $z_r$ in the following way: $x_e(z)=1$ for $z \ll z_r$ (possibly $x_e(z)=1.08$ or $x_e(z)=1.16$ for $z<3$ in order to take into account the first and second Helium ionization) and $x_e(z)<2\times 10^{-4}$ for $z>z_r$ in order to join the ionization fraction value after the recombination. In the following we will refer to this parametrization as ``sudden''  or ``instantaneous'' reionization. With this choice of parametrization there exist a one-to-one relation between the redshift of sudden reionization $z_r$ and the electron scattering optical depth $\tau_{e}$. The most recent constraints on the optical depth that come from the analysis of the Wilkinson Microwave Anisotropy Probe team on their seven-year data (WMAP7), in which it is assumed a sudden reionization scenario, is $\tau_{e}=0.088\pm0.015$. 
However, as already noticed, e.g. in \cite{mortonson},  and further emphasized by our previous works (\cite{Pandolfi1} and \cite{Pandolfi2}), the assumption of a general reionization scenario could affect the extraction of the constraints of cosmological parameters. In particular, we studied the effects of non-instantaneous reionization on the two principal inflationary parameters  (the scalar spectral index of primordial perturbations $n_s$ and the tensor-to-scalar ratio parameter $r$), and on the optical depth $\tau_{e}$.
The method used in the above cited works to describe a general reionization scenario, developed in Ref. \cite{mortonson}, is based on a principal components (PC) analysis of the reionization history, \xez. PCs provide a complete basis for describing the effects of reionization on large-scale $E$-mode polarization spectrum. 
Following Ref.\ \cite{mortonson}, one can treat \xez as a free function of redshift by decomposing it into its principal
components:
\begin{equation}
x_e(z)=x_e^f(z)+\sum_{\mu}m_{\mu}S_{\mu}(z),
\label{eq:xez}\end{equation}
where the principal components $S_{\mu}(z)$ are the eigenfunctions of the Fisher matrix describing the dependence of the polarization spectra on $x_e(z)$; the $m_{\mu}$ are the PC amplitudes for a particular reionization history, and $x_e^f(z)$ is the WMAP  {\it fiducial } model for which the Fisher matrix is computed and from which the PCs are obtained. 
Therefore the amplitude of eigenmode $\mu$ for a perturbation around the fiducial reionization history
$\delta x_e(z)\equiv x_{e}(z)-x_{e}^{f}(z)$ is
\begin{equation}
m_{\mu}=\frac{1}{z_{max}-z_{min}} 
  \int _{z_{min}}^{z_{max}} dz~S_{\mu}(z)\delta x_{e}(z).
\label{eq:xetommu}
\end{equation}

In Refs. \cite{Pandolfi1} and \cite{Pandolfi2} we made use of the publicly available $S_{\mu}(z)$ functions and varied the amplitudes $m_{\mu}$  for the first five eigenfunctions (i.e. for $\mu=1,...,5$). The principal components were computed only in the range of redshifts $z\in[6-30]$. 


In what follows we refer to this parametrization of reionization as the ``Principal Components''(PC) reionization.
Since the ionization fraction is bounded in $0 < x_e(z) < 1$ (neglecting helium reionization and the small residual ionized fraction after  recombination) in the range of redshifts in which PCs are defined, it is necessary to impose some limits on the amplitudes of the eigenmodes of equation (\ref{eq:xetommu}) to let the reionization fraction be within these limits, if only for the definition of reionization fraction. In Ref. \cite{mortonson} the authors find the ranges of values for the amplitudes $m_{\mu}$ compatible with $x_e(z) \in[0, 1]$ for all the redshifts in range of interest. In \cite{Pandolfi1} and \cite{Pandolfi2} we performed a Monte Carlo Markov Chains analysis  assuming a flat prior on (only) the ranges of values of the amplitudes $m_{\mu}$ whose linear combination with the function $S_{\mu}$ give a \xez in the allowed range. These values are reported in left part of Table \ref{tab:TableIV} and are labeled ``PC Bounds''.

However, these limits for the values of the PC amplitudes are a necessary but not sufficient condition for the reionization fraction to lie in $0<x_e(z)<1$. In fact, as noticed also by \cite{mortonson}, if any $m_{\mu}$ violates those bounds \xez is guaranteed to be unphysical in some redshift range, but the opposite is not true, because the full reionization history depends on the linear combinations of the product of the amplitudes times their corresponding PC principal component. Indeed, even if all the amplitudes $m_{\mu}$ satisfy the bounds reported in Table \ref{tab:TableIV}, $x_e(z)$ could assume an unphysical value for some redshifts. 
To overcome this potential problem, we have added in the version of the \texttt{cosmomc} package used in \cite{Pandolfi1} and \cite{Pandolfi2} the condition that the value of $x_e(z)$ computed at each step of a Markov Chain must be in the range $0 < x_e(z) < 1$ for every $z$. In these studies, this was the only ``physicality'' condition imposed on the possible reionization history. However, experimental data gathered in the last few years can be used to discard at least some of the possible \xez histories on well understood (astro)physical grounds. It is now possible to use reionization histories that are physically motivated and tested with known probes of the reionization epoch, such as the Gunn-Peterson optical depth, or the distribution in redshift of the Ly$\alpha$ emitters.

In this work we adopt the results of a well-tested semi-analytical reionization model proposed in Refs. \cite{CF2005} and \cite{CF2006} (in what follows we will refer to this model as the CF model). This model takes into account a large number of parameters and physical processes that are involved in modeling reionization, including (e.g.) the radiative and chemical feedbacks of the first sources of ionizing light on the evolution of the intergalactic medium (IGM), and constrain the model by comparing it with a variety of observational data, such as the redshift evolution of Lyman Limit Systems (LLS), the IGM temperature and the cosmic star formation density. Thus we will be able to build up an ensemble of reionization histories that is more robust from both the theoretical and the observational point of view, rather then rely on purely phenomenological, albeit model-independent, parameterization schemes as the PCs.

We will combine the CF model with a standard $\Lambda$CDM cosmological model and we perform a Monte Carlo Markov Chains analysis of the joint CMB and reionization data. We will thus be able to test the impact of considering a detailed physical model for reionization on the constraints of the cosmological parameters, and conversely to test the dependence of the CF model on the underlying cosmological model. 

At the end of such analysis we will moreover derive the subsequent constraints on the amplitudes of the reionization principal components $m_{\mu}$ (applying directly the equation \ref{eq:xetommu}).  By construction then, these limits on the values of amplitudes of the principal components will be compatible and constrained both by the CMB and by the astrophysical probes of the reionization process. 

The main objectives of the present work are then:
\begin{itemize}
	\item Verify the impact of considering a data-constrained and realistic reionization model on the determination cosmological parameters.
	\item Verify the impact on the constraints of the reionization parameters produced by variations of the cosmological parameters, i.e. refraining from fixing them a priori from the most updated best fit values of the WMAP experiment.
 	\item Obtain the PC amplitudes $m_{\mu}$ from the allowed reionization histories. 
 	\end{itemize}
 
As such an analysis with combined cosmological parameters characterizing the background evolution of the universe and astrophysical parameters modeling the reionization history has not yet been made, it is worthwhile to explore their mutual implications on the extraction of the constraints of the two ensemble of parameters.    

\begin{table*}[htbp]
		\begin{tabular}{l|c|c}
Parameter   &\ \ Mean\ \     &\  $95\%$\ c.l.\ limits\  \\
\hline
\hline
$\Omega_m$    & 0.2733       &[0.2260, 0.3305]\\
$\Omega_bh^2$ & 0.2184       &[0.0208, 0.0229]\\
$h$           & 0.6984       &[0.6553, 0.7422]\\
$n_s$         & 0.9579       &[0.9330, 0.9838]\\
$\sigma_8$    & 0.7941       &[0.7434, 0.8491]\\
\hline
$\epsilon_{II}$      & 0.0037       &[0.0016, 0.0067]\\
$\epsilon_{III}$     & 0.0165       &[0.0000, 0.0398]\\
$\lambda_{0}$ & 3.0152       &[1.0000, 5.1739]\\
\hline               
$\tau_{e}$		& 0.0803      & [0.0625, 0.1042]\\
$z_r$        &6.7469        & [5.8563, 8.2000]\\
		\end{tabular}
	\caption{Mean and 95\% c.l. constraints on the cosmological, astrophysical and derived parameters obtained with the reionization parametrized with the CF model of reionization.}
	\label{tab:TableI}
\end{table*}

\section{Analysis}
\label{sec:anal}
The details of the CF model are summarized in Ref. \cite{MCF2010}; in the present work we assume the following settings:

\begin{itemize}
\item We consider here a flat $\Lambda$CDM cosmology described by a set of cosmological parameters:
\begin{equation}
 \label{parameter}
      \{\Omega_m,\Omega_bh^2, h, \sigma_8, n_s\},
\end{equation}

where $\Omega_m$ is the total matter density relative to the critical density, $\Omega_bh^{2}$ is the baryonic matter density, $h$ is the reduced Hubble parameter $H_0=100 h$, $\sigma_8$ is the r.m.s. density fluctuation in spheres of radius $8h^{-1}$ Mpc and $n_s$ is the scalar spectral index of primordial perturbations. We want to stress that these cosmological parameters are considered here as free parameters, so that they are not assumed a priori, as in \cite{MCF2010}.

\item The CF reionization model contains additional three free parameters. These are
$\epsilon_{\rm II, III} = [\epsilon_* f_{\rm esc}]_{\rm II,III}$, the product of the star-forming efficiency (fraction of baryons within collapsed haloes going into stars) $\epsilon_*$ and the fraction of photons escaping into the IGM $f_{\rm esc}$  for PopII and PopIII stars; 
the normalization $\lambda_0$ of the  ionizing photons mean free path (see Ref. \cite{MCF2010} for details). In what follows we refer to these three parameters as the ``astrophysical'' parameters, to distinguish them from the five ``cosmological'' ones described described in the previous point. 

\item  The ranges of variation adopted for the three free astrophysical parameters are $\epsilon_{\rm II} \in [0; 0.02]$, $\epsilon_{\rm III} \in [0; 0.1]$, $\lambda_0 \in [1;10]$.

\item  The observational data used to compute the likelihood analysis are (i) the photo-ionization rates $\Gamma_{\rm PI}$ obtained using Ly$\alpha$ forest Gunn-Peterson optical depth observations and a large set of hydrodynamical simulations \cite{Bolton} and (ii) the redshift distribution of LLS $\de N_{\rm LL}/\de z$  in the redshift range of $0.36 < z < 6$ \cite{Songaila}.
The data points are obtained using a large sample of QSO spectra. For details, see Ref.\ \cite{MCF2011}.

\item In order to make the analysis self-consistent, the WMAP7 constraint on the total electron scattering optical depth $\tau_{e}$ is not considered in this analysis. This prevents a possible loophole in our analysis: WMAP7 constraints on $\tau_{e}$ have been  obtained using the assumption of instantaneous reionization at $z=z_r$. Once this idealized evolution of \xez is dropped (this paper), the value of  $\tau_{e}$ must be a byproduct of the new analysis rather than being inserted artificially as an external constraint into it.  Moreover, as already pointed out in Ref. \cite{MCF2010}, the CMB polarization spectra are sensitive to the shape of the reionization history and considering a more general reionization scenario could lead to a tighter optical depth constraint than derived by WMAP7 \cite{Pandolfi1}.

\item Finally, we impose the prior that  reionization should be completed by $z=5.8$ to match the flux data of Ly$\alpha$ and Ly$\beta$ forest.

\end{itemize}

With these hypotheses we have then modified the Boltzmann CAMB code \cite{camb} to incorporate the CF model and performed a MCMC analysis based on an adapted version of the public available MCMC package \textsc{cosmomc} \cite{Lewis:2002ah}. Our basic data set is the seven--yr WMAP data \cite{wmap7} (temperature and polarization), on top of which we add two ``astrophysical'' datasets, i.e. the LLS redshift evolution, $dN_{LL}/dz$  Ref. \cite{Songaila}, and the Gunn-Peterson optical depth measurements presented in Ref.  \cite{Bolton}.  To extract the constraints on free parameters from such combined data set we consider a total likelihood function $L \propto \exp (-\mathcal{L})$ made up by two parts: 
\begin{equation}
{\mathcal L} = \f{1}{2}\sum_{\alpha=1}^{N_{\rm obs}} \left[
\f{{\cal J}_{\alpha}^{\rm obs} - {\cal J}_{\alpha}^{\rm th}}{\sigma_{\alpha}}
\right]^2 + {\cal L'}
\end{equation}
where $\mathcal{L}'$ refers to the WMAP7 likelihood function and is computed using the routine supplied by the WMAP team; ${\cal J}_{\alpha}$ represents the set of $N_{\rm obs}$ observational points referring to Gunn-Peterson optical depth LLS distribution data; finally, $\sigma_{\alpha}$ are the corresponding observational error-bars. We constrain the free parameters by maximizing ${\mathcal L}$ with flat priors on the allowed parameter ranges and the aforementioned prior on the end of reionization at $z=5.8$.

The Monte Carlo-Markov Chain convergence diagnostics are done on 4 chains applying the Gelman and Rubin ``variance of chain mean''/``mean of chain variances'' $R$
statistic for each parameter. We considered the chains to be converged at $R-1<0.03$.

\begin{table*}[htbp]
		\begin{tabular}{l|c|c|c}
Parameter     &\ \ WMAP7\ \ &\ \ WMAP7 + PC &\ WMAP7\ +\ ASTRO    \\
\hline
\hline
$\Omega_m$    & $0.266\pm0.029$     &  $0.243\pm0.032$  &$0.273\pm0.027$   \\
$\Omega_bh^2$ & $0.02258^{+0.00057}_{-0.00056}$   & $0.02321\pm0.00076$ & $0.02183\pm 0.00054$     \\  
$h$           & $0.710\pm0.025$     & $0.735\pm 0.033$ & $0.698\pm0.023$        \\ 
$n_s$         & $0.963\pm0.014$     & $0.994\pm0.023$ &$0.958\pm 0.013$           \\  
$\sigma_8$    & $0.801\pm0.030$     & ----- &$0.794\pm 0.027$                \\  
\hline               
$\tau_{e}$		& $0.088\pm0.015$    & $0.093\pm0.010$  & $0.080 \pm 0.012$             \\ 
$z_r$*         & $10.5\pm 1.2$      & -----  & $6.7\pm 0.6$          \\ 
		\end{tabular}
	\caption{Comparison of the 68\% c.l. posterior probabilty contraints obtained for different parametrizations of reionization.\\
	* The $z_r$ parameter has a different definition in the different reionization scenarios (see text for details).}
	\label{tab:TableII}
\end{table*}

\begin{table*}[htbp]
		\begin{tabular}{l|c|c|c|c}
Parameter     &WMAP7 + ASTRO Mean  & WMAP7 + ASTRO $95\%$ c.l. limits &    CF Mean  & CF $95\%$\ c.l.\ limits\  \\
\hline
\hline
$\epsilon_{II}$      & 0.0037       &[0.0016, 0.0067] & 0.003       &[0.001, 0.005]\\
$\epsilon_{III}$     & 0.0165       &[0.0000, 0.0398] & 0.020       &[0.0000, 0.043] \\
$\lambda_{0}$ & 3.0152       &[1.0000, 5.1739] & 5.310       &[2.317, 9.474]\\
\hline               
$\tau_{el}$		& 0.0803      & [0.0625, 0.1042]&  $\equiv0.088\pm0.015$   &    $\equiv0.088\pm0.015$              \\
$z_r$         & 6.7469        &[5.8563, 8.2000]& 6.762        & [5.800, 7.819]\\
		\end{tabular}
	\caption{Comparison between the mean value and the 95\% c.l posterior constraints between the present work (WMAP7 + ASTRO) and the CF model, Ref. \cite{MCF2010} (MCF). }
	\label{tab:TableIII}
\end{table*}
                  
\section{Results}
\label{Res}
The results of the MCMC analysis described above are summarized in Table \ref{tab:TableI}, where we list the marginalized posterior probabilities at 95\% confidence level (c.l.) errors on the free cosmological and astrophysical parameters. We also report the constraints for two derived parameters: the electron scattering optical depth $\tau_{e}$ and reionization redshift $z_r$, to be intended as the redshift at which the reionization is 99\% complete.  
In Table \ref{tab:TableII} we show the 68\% c.l. constraints obtained by the WMAP team for the standard 6-parameter $\Lambda$CDM model (``WMAP7'') and the constraints obtained on the cosmological parameters from the present analysis (``WMAP7 + ASTRO'').

As we can see from the Table \ref{tab:TableII} the results of our work mildly differ from the WMAP7 results for the parameters of the standard $\Lambda$CDM model. The most sensitive parameter for the presence of the ``astrophysical'' datasets (LLS and Gunn-Peterson data)  is $\Omega_bh^2$ whose mean values in the two cases differ by more than a standard deviation from each other. It is important to note that even when considering a complex reionization history implying three new parameters the errors remain practically the same as in the standard case.

Table \ref{tab:TableII} reports the results obtained in \cite{Pandolfi1} for the WMAP7 dataset with the PC reionization (``WMAP7 + PC''). 
This method produces two main differences with respect to the WMAP7 + ASTRO case: the first is related to the constraints obtained for $n_s$. In \cite{Pandolfi1} the constraints for the scalar spectral index were compatible with $n_s=1$, i.e. the Harrison-Zel'dovich (HZ) primordial power spectrum, when instead WMAP7+ASTRO excludes the value $n_s=1$ at  $>3\sigma$. The second difference concerns $\tau_{e}$ in the two cases: for WMAP7 + PC this quantity is in the range  $\tau_{e}=0.093\pm0.010$, while the  WMAP+ASTRO case gives a mean value lower by $>1-\sigma$, i.e. $\tau_{e}=0.080\pm0.012$.
Note that in the WMAP7 + PC case we did not consider constraints on the $\sigma_8$ parameter, so in Table \ref{tab:TableII} the corresponding value is missing.

There is a caveat in comparing the constraints obtained on $z_r$. Indeed, in the WMAP7 case $z_r$ is the redshift at which the universe undergoes an instantaneous and complete reionization process. In the more realistic, extended reionization scenarios considered here instead, $z_r$  is defined as the redshift at which the IGM is 99\% re-ionized by volume. With this clarification in mind, WMAP7+ASTRO results predict $5.8 <z_r < 8.2$ at 95\% c.l. (see Table \ref{tab:TableI}). 

\begin{figure}[htbp]
	\centering
		\includegraphics[width=6cm, angle=-90]{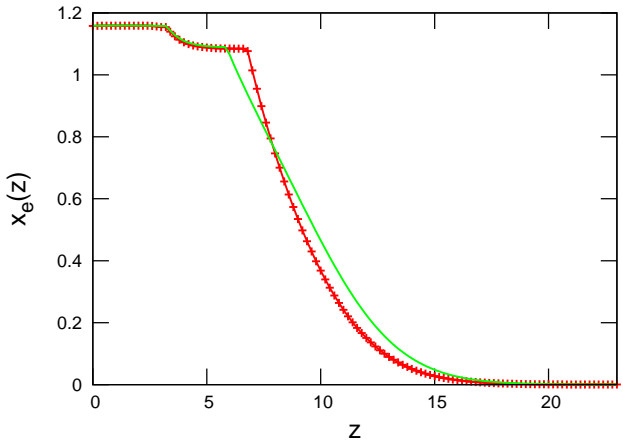}
	\caption{Ionization histories for the best-fit model for the two cases WMAP7+ASTRO (red dotted solid curve) and CF (green solid curve) \cite{MCF2010}.}
	\label{bf}
\end{figure}

In Table \ref{tab:TableIII} we report the 95\% c.l. posterior probability constraints for the reionization parameters $\epsilon_{II}$, $\epsilon_{III}$ and $\lambda_0$ obtained in the present work (WMAP7 + ASTRO case, cosmological parameters free to vary) compared to those obtained in Ref. \cite{MCF2010} in which the cosmological parameters were fixed to the WMAP7 best fit values (CF case).  Figure \ref{bf} shows the comparison between the best-fit model for the \xez evolution for the two cases of WMAP7 + ASTRO and CF. For the WMAP7 + ASTRO case, full hydrogen reionization is not only achieved earlier than in the CF model, but the evolution is faster, resulting in an initially
lower \xez above $z=8$. These differences are entirely induced by the fact that we have now allowed the cosmological parameters to vary together with the astrophysical ones, but they are relatively small. The fact that the astrophysical parameters do not show much dependence on cosmology is understandable because the cosmological parameters affect the  reionization process mostly through structure formation. The next obvious step is to include large scale structure information in the analysis. In conclusion, including astrophysical datasets in the analysis seems to lead to relatively important effects on the extraction of the cosmological parameters.

\subsection{PC amplitude reconstruction}

\begin{table*}[htbp]
		\begin{tabular}{lcc}
Parameter & PC Bounds &   Astrophysical Bounds\\
\hline
\hline
$m_{1}$  &$[-0.1236, 0.7003]$& $[-0.1229, -0.0866]$ \\
$m_{2}$  &$[-0.6165, 0.2689]$& $[-0.2594, 0.0002]$\\
$m_{3}$  &$[-0.3713, 0.5179]$& $[0.0763,\ 0.2941]$\\ 
$m_{4}$  &$[-0.4729, 0.3817]$& $[-0.2107,  -0.1080]$\\
$m_{5}$  &$[-0.3854, 0.4257]$& $[0.0418,\ 0.1319]$\\
\hline               
 		\end{tabular}
	\caption{Ranges of variation for the amplitudes of the principal component, in the case of the Principal Components and in the case of the 99\% c.l. reconstructed amplitudes of the present analysis (see text for details).}
	\label{tab:TableIV}
\end{table*}

For each reionization history allowed by the MCMC likelihood analysis, we use eq. (\ref{eq:xetommu}) to reconstruct the amplitudes of the first five PC amplitudes, $m_{\mu}$, with $\mu=1...5$. By construction now, the amplitudes $m_{\mu}$ not only fulfill the necessary physicality conditions (see Sec. 1) but also they are compatible with the additional astrophysical data sets considered in this analysis, i.e. the Ly$\alpha$ Gunn-Peterson test and the LLS redshift distribution.

\begin{figure}[htbp]
	\centering
\includegraphics[width=5cm]{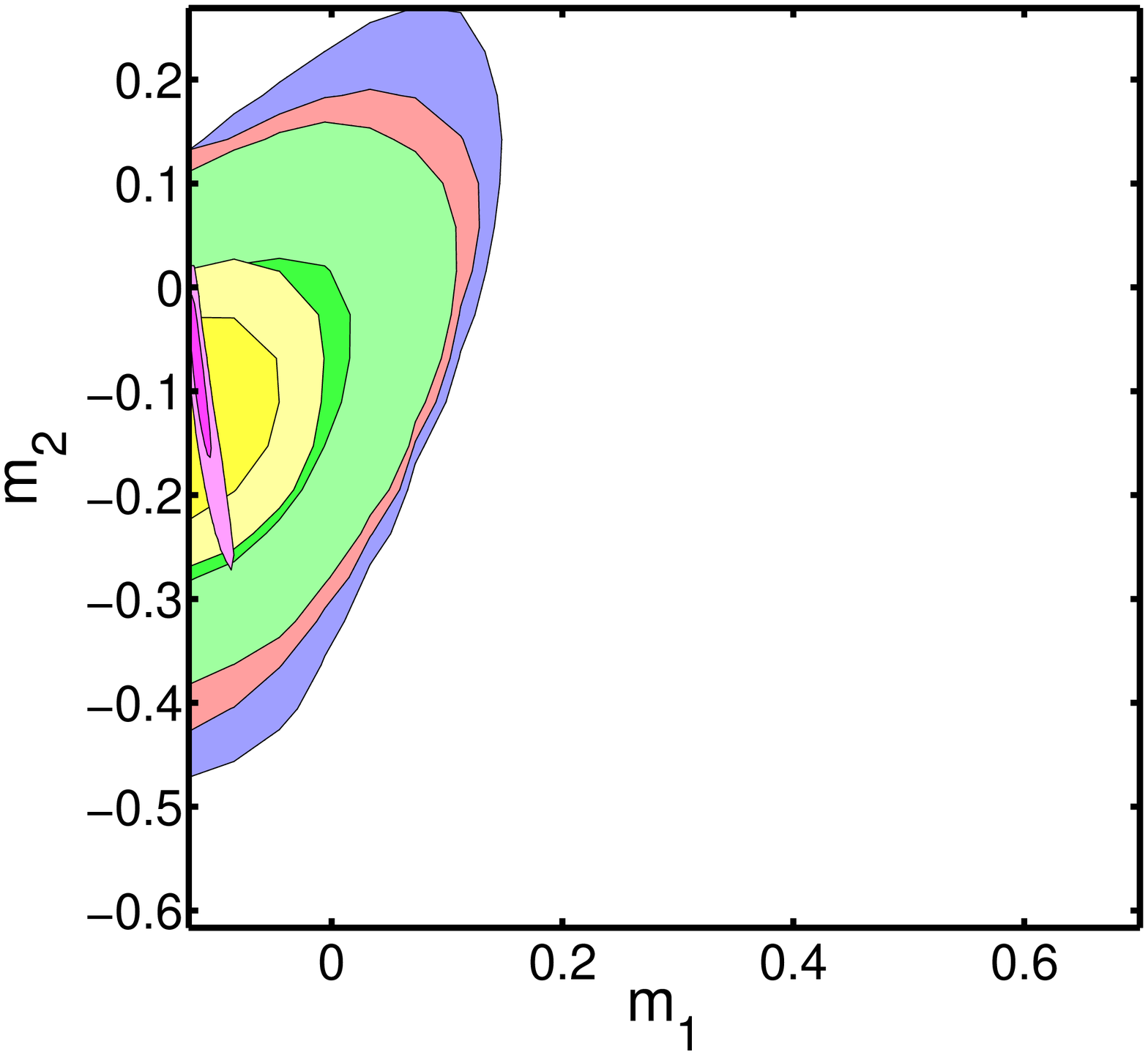}
\includegraphics[width=5cm]{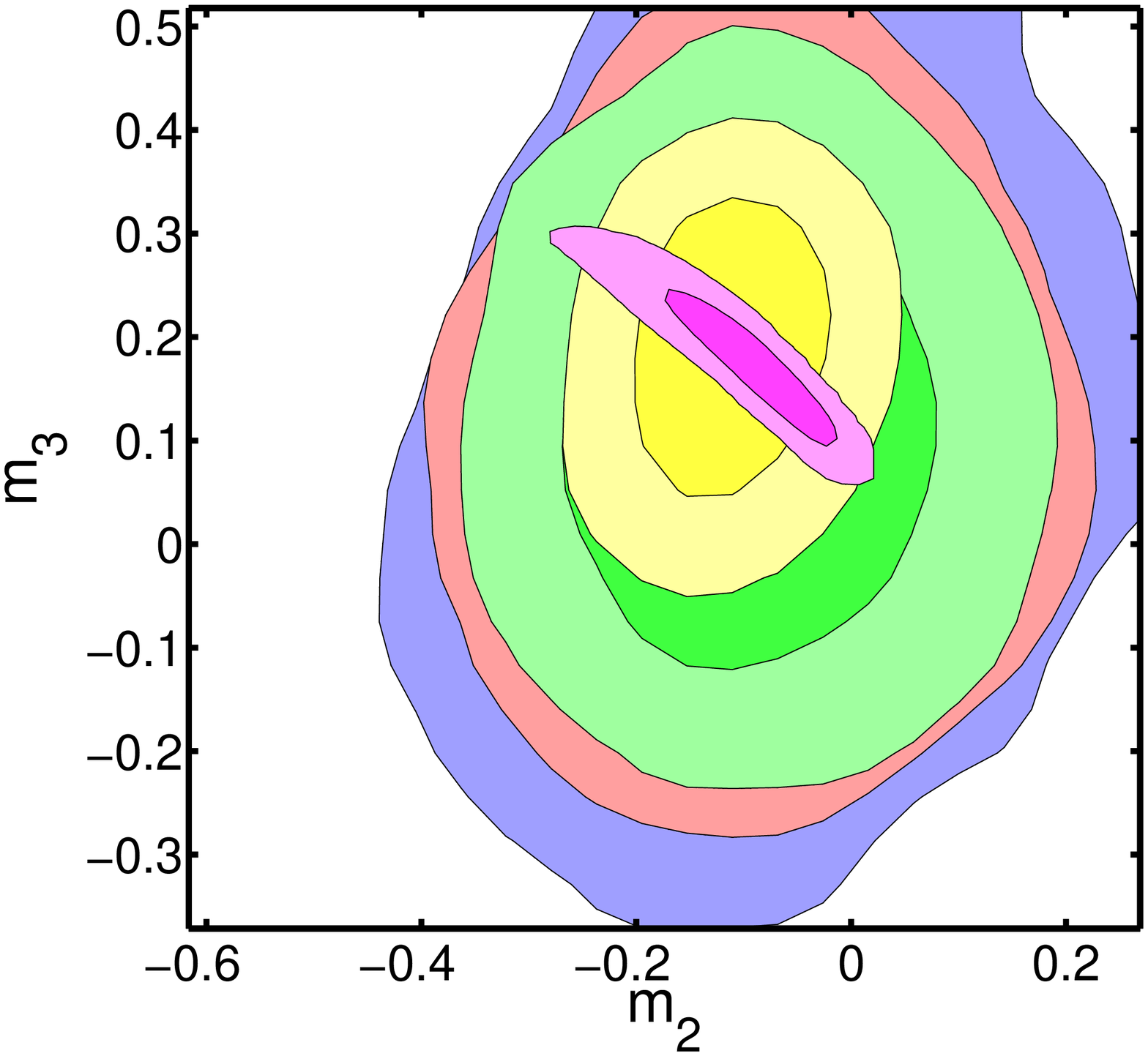}
\includegraphics[width=5cm]{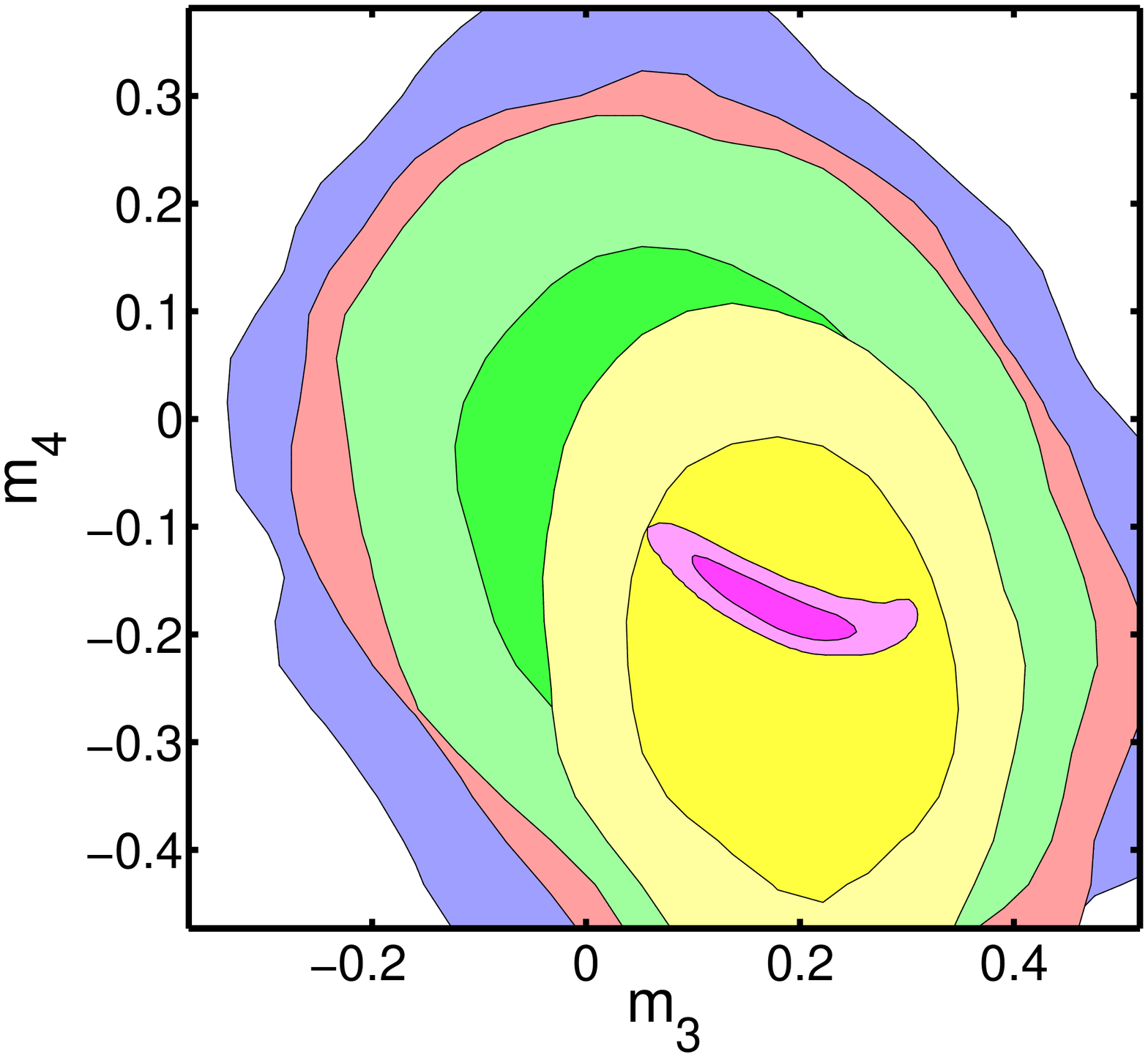}
\includegraphics[width=5cm]{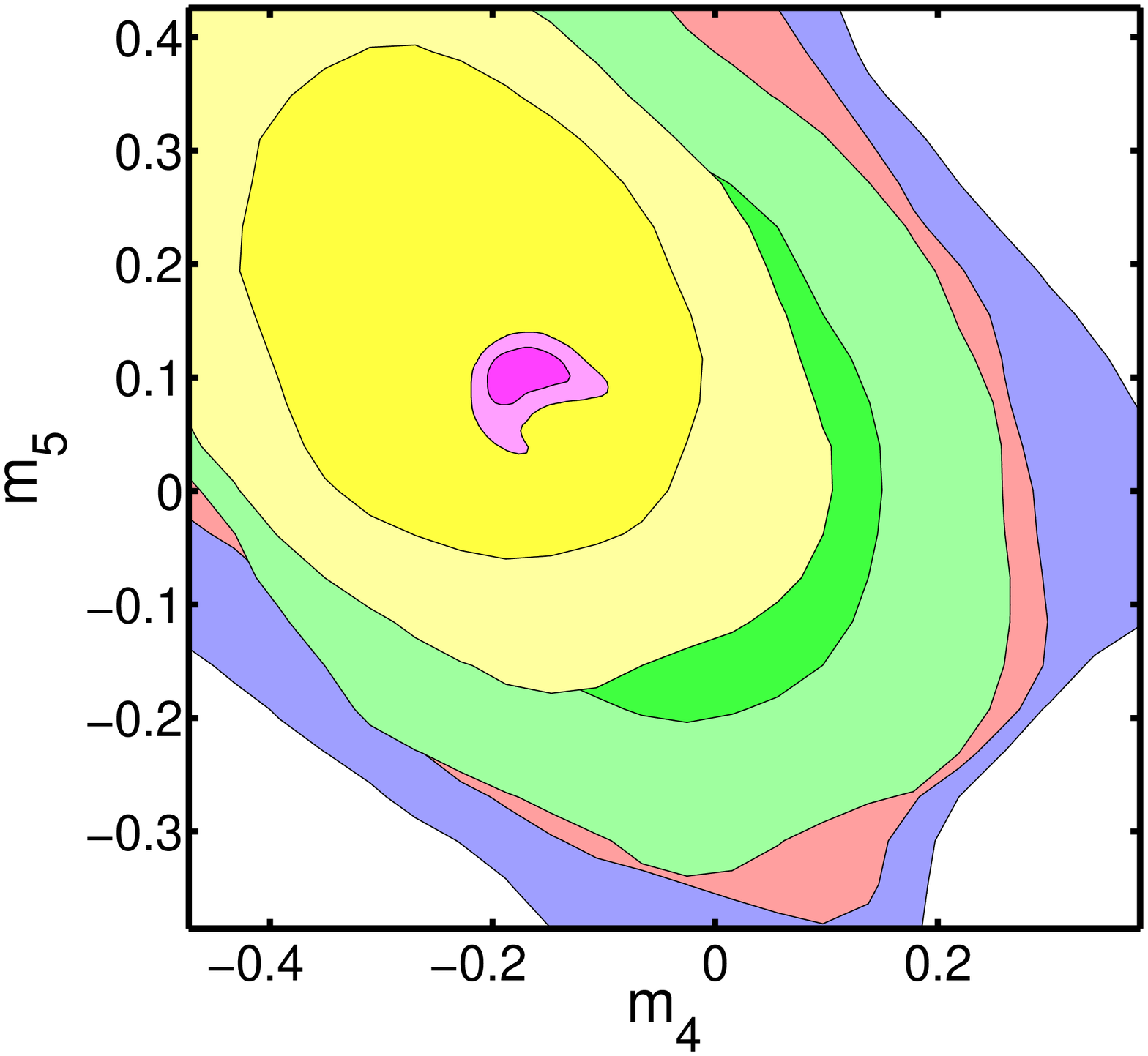}
	 \caption{68\% and 99\% reconstructed c..l. constraints for the values of the PC amplitudes computed from CF model and eq. \ref{eq:xetommu} (top layer, pink). Background contours refer to  68\% and 95\% c.l. constraints obtained in \cite{Pandolfi1} with the PC reionization for WMAP7 (bottom layer, blue), WMAP7+QUAD+ACBAR+BICEP (CMBAll, next layer up, red), CMB All + LRG-7 (next layer, green) and simulated Planck data (next layer, yellow), respectively.}
	\label{m_i}
\end{figure}

In Fig. \ref{m_i}  we show the two dimensional  68\% and 99\% c.l constraints for the amplitudes $m_{\mu}$ obtained here compared with those obtained in \cite{Pandolfi1} for which we show the two dimensional 68\% and 95\% c.l. distributions for each of the cases considered. 
We choose to report the 99\% c.l. instead of the usual 95\% c.l. limits to be as conservative as possible in showing the reionization histories allowed by the MCMC likelihood analysis.  The color (layer) code is the following: in pink (top layer) there is the case WMAP7 + ASTRO considered in the present work. In the background there are the cases considered in \cite{Pandolfi1}: in blue is the WMAP7 case (bottom layer) , in red (next layer up) is the case called ``CMB All'' ( i.e. WMAP7 + ACBAR + BICEP+ QUAD + BOOMERanG), green (next layer) is CMB All + LRG-7 and yellow (next layer) is simulated Planck data. \cite{Pandolfi1} considered an ensable of CMB dataset along with WMAP7, and also we forecasted future constraints from the Planck experiment, simulating a set of mock data with a fiducial model given by the best fit WMAP5 model with the following experimental noise: 
\begin{equation}
N_{\ell} = \left(\frac{w^{-1/2}}{\mu{\rm K\mbox{-}rad}}\right)^2
\exp\left[\frac{\ell(\ell+1)(\theta_{\rm FWHM}/{\rm rad})^2}{8\ln 2}\right],
\end{equation}
\noindent where $w^{-1/2}$ is the temperature noise level (a factor $\sqrt{2}$ larger for polarization noise) and  $\theta$ is the beam size.
For the Planck mission we use $w^{1/2}=58 \mu K$ and $\theta_{\rm FWHM}=7.1'$ equivalent to expected sensitivity of the $143$ GHz channel.

The region spanned by PC amplitude values is much smaller than that allowed by when the PC bounds only are imposed.  The 99\% c.l. constraints values are reported in the right part of the Table (\ref{tab:TableIV}) (``Astrophysical Bounds'').  As seen from Table (\ref{tab:TableIV}) the amplitudes of all the principal components (except for $m_2$) obtained with the above procedure are constrained at 99\% c.l. to take a definite sign, negative for $m_1$ and $m_4$ and positive for $m_3$ and $m_5$. Moreover, even if the 99\% c.l. upper bound of $m_2$ is positive, this second amplitude is mostly constrained to be always negative.  These results are in qualitative agreement with \cite{Pandolfi1}, who also found that the same amplitude signature, albeit with errors large enough that the 95\% c.l. bounds encompass values of both possibile signs.

\section{CONCLUSIONS}
\label{sec:concl}
With the aim of constraining the evolution of cosmic reionization, we have extended previous work based on the use
of Principal Components analysis. The main novelty of the present work is represented on one hand by complementing
available CMB data with additional astrophysical results from quasar absorption line experiments, as the Gunn-Peterson 
test and the redshift evolution of Lyman Limit Systems. In addition, we have for the first time explored the effects of a 
joint variation of both the cosmological ($\Omega_m,\Omega_bh^2, h, \sigma_8, n_s$) and astrophysical ($\epsilon_{\rm II}, 
\epsilon_{\rm III}, \lambda_0$, see Sec \ref{sec:anal} for their physical meaning) parameters. Note that, differently from the vastly
used approach in the literature, we do not impose a priori any bound on the electron scattering optical depth $\tau_e$,
which instead we calculate a posteriori. This is to prevent a possible loophole in the calculation, as the WMAP determination
of such quantity is based on the assumption of an instantaneous reionization which we do not make here.  

Including a realistic (i.e physically motivated) reionization history in the analysis induces mild changes in the 
cosmological parameter values deduced through a standard WMAP7 analysis. Particularly noteworthy are the 
variations in  $\Omega_bh^2 = 0.02258^{+0.00057}_{-0.00056}$ (WMAP7)   vs.  $\Omega_bh^2 = 0.02183\pm 0.00054$ (WMAP7 + ASTRO),
and the new constraints for the scalar spectral index, for which WMAP7+ASTRO excludes the Harrison-Zel'dovich value $n_s=1$ at $>3\sigma$.
Finally, the e.s. optical depth values is considerably decreased with respect to the standard WMAP7, i.e. $\tau_{e}=0.080\pm0.012$.
We conclude that inclusion of astrophysical datasets, allowing to robustly constrain the reionization history,  in the extraction procedure 
of cosmological parameters leads to relatively important differences in the final determination of their values.

\section*{AKNOWLEDGMENTS}
SP is grateful for the hospitality and support during his research at Scuola Normale Superiore in Pisa in
March-July 2011. Support was given by the Italian Space Agency through the ASI contracts Euclid-IC (I/031/10/0).


\end{document}